\documentclass[acmsmall]{acmart}
\AtBeginDocument{%
  \providecommand\BibTeX{{%
    \normalfont B\kern-0.5em{\scshape i\kern-0.25em b}\kern-0.8em\TeX}}}

\setcopyright{acmlicensed}
\acmJournal{PACMHCI}
\acmYear{2022} \acmVolume{6} \acmNumber{CSCW2}
\acmArticle{453} \acmMonth{11} \acmPrice{15.00}
\acmDOI{10.1145/3555554}

\usepackage{todonotes}
\newcommand\revision[1]{\textcolor{black}{#1}}
\newcommand\rev[1]{\textcolor{black}{#1}}

\begin{document}

\title[Organizational Distance Also Matters]{Organizational Distance Also Matters: How Organizational Distance Among Industrial \revision{Research} Teams Affect Their \revision{Research} Productivity}

\author{Dakuo Wang}
\email{dakuo.wang@ibm.com}
\affiliation{%
  \institution{IBM Research, }
  \institution{Northeastern University}
  \country{USA}
}

\author{Michael Muller}
\affiliation{%
  \institution{IBM Research}
  \country{USA}
}

\author{Qian Yang}
\affiliation{%
  \institution{Cornell University}
  \country{USA}
}

\author{Zijun Wang}
\affiliation{%
  \institution{IBM Research}
  \country{USA}
}

\author{Ming Tan}
\affiliation{%
  \institution{IBM}
  \country{USA}
}

\author{Stacy Hobson}
\affiliation{%
  \institution{IBM Research}
  \country{USA}
}

\renewcommand{\shortauthors}{Wang, et al.}

\begin{abstract}
Geographically distributed teams often face challenges in coordination and collaboration, lowering their productivity. Understanding the relationship between team dispersion and productivity is critical for supporting such teams. Extensive prior research has studied these relations in lab settings or using qualitative measures. This paper extends prior work by contributing an empirical case study in a real-world organization, using quantitative measures. We studied 117 new research project teams from the same discipline within an industrial research lab for 6 months. During this time, all teams shared one goal: submitting research papers to the same target conference. We analyzed these teams' dispersion-related characteristics as well as team productivity. Interestingly, we found little statistical evidence that geographic and time differences relate to team productivity. However, organizational and functional distances are  predictive of the productivity of the dispersed teams we studied. We discuss the open research questions these findings revealed and their implications for future research.
\end{abstract}



\maketitle
\section{Introduction}

\rev{Distributed} teams -- teams in which people work toward a shared goal from more-than-one geographic locations -- are becoming increasingly common in companies and industries~\cite{team-dispersion-Bardhan}.
{The geographically distributed characteristic is only one facet of today's dispersed team.}
\rev{Prior research has identified many dimensions of dispersion beyond geographically distributed team characteristic that have an impact on team productivity
\cite{o2010go,team-dispersion-Bardhan,van2007work,muller2019learning,olson2000dnflustance,yim2017synchronous}}
For example, \textit{functional} distance (team members having different expertise and job roles) and \textit{organizational} distance (hierarchical and peer relations among team members) could both negatively influence teamwork, \rev{as the communication barrier is higher across different expertise and different organizational units~\cite{olson2000dnflustance}}. And on top of that \textit{temporal} and \textit{geographical} distances could substantially exacerbate these influences \cite{harrison1996re}. 
These dimensions, among others, together can add burdens to communication, coordination, and collaboration, thereby hindering productivity \cite{olson2000dnflustance,define-distributed-teams-Maznevski2000}. 

Examining the relationship between various dimensions of team dispersion\footnote{\rev{In this paper, we use team dispersion and team distance interchangeably.}} and team productivity is critical for understanding and supporting dispersed teamwork~\cite{grudin1995groupware,wen2017supporting}.
Olson and Olson have provided an influential list of these dimensions in 2000~\cite{olson2000dnflustance}, which has since become a fruitful line of Computer Supported Cooperative Work (CSCW) research (e.g., \cite{ahmed2012distance, bjorn2014does, bradner2002distance, chan2014conceptual,williams1998demography,hinds2003out,lambiase2022good,yim2017synchronous,olson2017people}).

Despite this rich body of work, establishing relationships between team dispersion and productivity remains challenging in real-world organizational contexts.
It requires researchers to closely track multiple comparable teams that are working toward similar goals. 
\rev{Collecting such a dataset is not easy in real-world settings, where project teams typically take shape at different times, work towards different goals, adhere to their respective timelines, and the dataset may not be accessible to researchers.}
As a result, \rev{CSCW} researchers more often studied temporarily-formed teams in lab settings~\cite{yim2017synchronous,wang2015docuviz}, or self-organized teams on online crowd-sourcing or MOOC platforms instead (e.g., ~\cite{wen2017supporting,zheng2015impact}). 
However, the dispersion-related relationship within these temporarily-formed teams may be different from those in real-world organizations.
When researchers did gain access to observe real-world teams, \rev{many measured proxies of team productivity (such as the effectiveness of within-team communication) rather than team outcome itself \cite{olson2000dnflustance,gergle2014experimental}.}

Building upon and extending prior work, we wanted to \textit{quantitatively} examine the relationship between team dispersion and productivity in a real-world organizational setting.
Specifically, we ask: \textbf{Whether and how are the following quantitative measures of team dispersion related to team productivity? }

\begin{itemize}
    \item Geographic \rev{distance} (\revision{whether team members share the same office location})
    \item Temporal \rev{distance} (how far away are each pairs of team members in terms of the time zone they work from)
    \item Functional \rev{distance} (whether team members share the same job roles e.g. engineers, researchers, managers) 
    \item Organizational \rev{distance} (how far away are each pairs of team members on the organizational hierarchy chart)
\end{itemize}

In 2018, we encountered a rare opportunity to investigate these questions: we gained access to a large, multinational company to study its 117 newly-formed research teams. These teams consisted of employees of different job roles, working from different geographic locations and time zones. All shared the same goal of submitting a publication to the same prestigious research conference. All had six months to accomplish this goal until the paper deadline arrives.
The fact that these teams differed in the aforementioned dispersion-related characteristics, yet all shared the same clearly defined and measurable/observable goal (i.e. paper submission) uniquely allowed us to study the relationships between team dispersion and productivity.


This paper starts with a review of prior examinations of these relationships, most of which were based on controlled lab studies, studies of self-organized online teams, and qualitative on-site studies. We then report the data we collected from our research site and the quantitative dispersion and productivity measures we derived. Our analysis did \textit{not} reveal significant effects of temporal or geographical dispersion on team productivity; this can seem at tension with the common assumption that the most thorny challenge of geographically dispersed teamwork is geographic differences \cite{team-dispersion-Bardhan}. In contrast, less organizational dispersion of a team is significantly related to (and in fact, predicts) higher productivity. We discuss how our results echo or contradict prior research findings, identifying issues that merit further research as well as implications for supporting dispersed teamwork in organizations.

This paper makes two major contributions: First, it synthesizes prior work and identifies a set of quantitative measures of team dispersion characteristics. This provides a first step towards systematically and quantitatively examining team dispersion characteristics. 
Second, this paper offers a rare quantitative analysis of how characteristics of team dispersion impact team productivity in a real-world organizational setting. This surfaced intriguing insights, some of which can seem at tension with findings from prior qualitative research, therefore, merit more explicit debate and discussion.

\section{Related Work}

To contextualize our work, we first review the dimensions of team dispersion and the ways in which prior work quantified them. Our study design heavily referenced these quantification methods.
We then describe the relationship between team dispersion and productivity according to existing literature.

\subsection{Measuring Dimensions of Team Dispersion}\label{related-work-measures}

\textit{Teams} are groups of people who are tasked with a shared goal.
\textit{Geographically dispersed teams} -- also known as distributed or virtual teams -- are teams in which members come from different geographic locations, usually with heavy reliance on computer-mediated communication \cite{define-distributed-teams-Maznevski2000}. 
Various characteristics of dispersed teams can impact their productivity \cite{team-dispersion-Bardhan, cramton2004subgroup, griffith2003virtualness, o2010go, Polzer-2006-faultline}. These characteristics include and are limited to:

\begin{itemize}
    \item \textit{Geographic distances} among team members. Geographical distance (proximity) is the physical distance between team members. \revision{Studies have shown that employees working 30 meters apart are similar to the employees working remotely~\cite{kraut1995coordination,allen1984managing}. Thus, in our paper we approximate the team members as collocated if they work in the same building, otherwise we consider them as distributed.}
    \item \textit{Temporal distances} among team members.
    Temporal distance is a closely related to geographical distances among team members. Teams distributed across time zones could face additional challenges of communication and coordination~\cite{olson2000dnflustance};
    
    \item \textit{Functional distances} between team members:
    Functional distance refers the extent to which team members share the expertise and serve the same job roles (e.g. researchers, designers, etc.).
    Functional distance between team members affect how they share knowledge and expertise, thereby influencing teamwork.
    
    \item \textit{Socio-demographic differences} among team members: This includes both differences in personal traits and demographic backgrounds. For example, gender and age composition \cite{abramo2013gender, rogelberg1996gender}, personal traits \cite{lykourentzou2016personality}, cultural background \cite{kayan2006cultural}, social media usage of team members~\cite{gomez2019would, horwitz2007effects} are known to have a considerable impact on teamwork. 
    
    \item \textit{Organizational distance} among team members: That is, how far away each pairs of team members are on their organizations' hierarchy chart, a diagram that lays out the reporting structure (who reports to whom) as well as peer relations (who report to the same manager) within the organization (Figure \ref{org-chart-demo}). 
    
    \begin{figure}[h!]
        \includegraphics[scale=0.5]{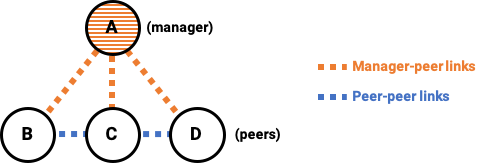}
        \centering
        \captionsetup{width=0.9\textwidth}
        \caption{An example organizational hierarchy chart. It annotates hierarchical and peer relations among employees.}
        \label{org-chart-demo}
    \end{figure}
    
    We noticed two different ways of conceptualizing organizational relations in prior research.
    First is a strict hierarchical (SH) approach, which considers the organization as a strictly hierarchical network \cite{kolari2007structure}. This approach emphasizes the hierarchical/power relations and does \textit{not} consider links between peer team members. 
    The second approach is a strict hierarchical with peers (SHP) approach, which considers organization primarily as a social network \cite{zhang2005searching}. This approach highlights the importance of peer relations in teamwork, in addition to hierarchical ones.
    
    These two ways of conceptualizing organizational relations lead to two ways of quantifying organizational distance.
    Consider a team of manager A and employees B, C, and D. Figure \ref{SHvSHP} illustrates its organizational hierarchy chart. Through the SH lens, the organizational distance between each pair of employees is 2 (highlighted in gray in Figure \ref{SHvSHP} left). In contrast, the SHP approach also considers peer relations, therefore the organizational distance between each pair of employees is 1 (Figure \ref{SHvSHP} right).
    These different approaches in quantifying organizational distance have great implications for organizational research, for example, in understanding how technologies can enable information to most effectively flow across the organization, thereby facilitating teamwork.
    
    \rev{As an example, in an organization chart, everyone is connected with everyone else. So if a \textit{project} team has two members from the same \textit{organizational} team, the project team's organizational distance will be 1; if the new \textit{project} team has two members from two different \textit{organizational} teams, but their managers report to the same upper manager, then the new project team's organizational distance will be 3.}
    
    \begin{figure}[h!]
        \includegraphics[scale=0.5]{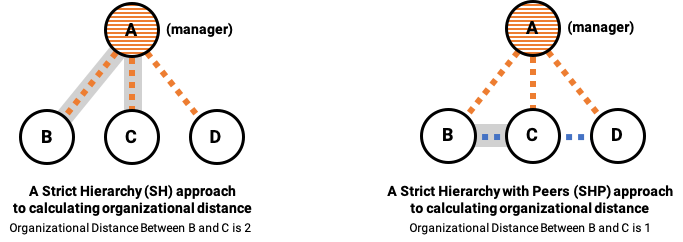}
        \centering
        \captionsetup{width=0.9\textwidth}
        \caption{The difference between a strict hierarchical (SH) approach \cite{zhang2005searching} and a strict hierarchical with peers (SHP) approach \cite{kolari2007structure} to calculating organizational distance between team members. }
        \label{SHvSHP}
    \end{figure}
    

\end{itemize}


\subsection{Dimensions of Team Dispersion and Productivity}

Decade-long research has investigated how each dimension of team dispersion impacts team productivity.
Consensus has started to emerge on some of these dimensions. For example with regard to \textit{socio-demographic differences}, studies have repeatedly shown that demographically-diverse and personality-wise-balanced teams are more productive than the less diverse or less balanced ones \cite{lykourentzou2016personality,gomez2019would,layton2007software}; that ``\textit{diversity improves creativity}" \cite{anagnostopoulos2012online, lykourentzou2016personality}.
On other team-dispersion dimensions, prior research has produced mixed results.
For example with regard to \textit{organizational distance}, some have argued that peer relations are stronger than manager-peer relations therefore can improve team outcomes \cite{muller2016influences, mitra2017spread}, while other studies showing that strong manager-peer relations reduce delays and team member conflicts \cite{hinds2015flow, grinter1999geography, herbsleb2000distance}. 
A more comprehensive review of this body of research can be found elsewhere~\cite{olson2000dnflustance,harris2019joining}. 

O'Leary and Cummings' landmark work, in particular, argued for conceptualizing team dispersion as a ``\emph{multidimensional construct}''; that each dimension is theoretically linked with different team outcomes and they simultaneously influence a team's productivity\cite{o2007spatial}.
Among these dimensions, prior research suggested that geographical and temporal dimensions are particularly important \cite{harrison1996re, olson1993groupwork,olson2000dnflustance}. Empirical studies showed that dispersed teams often devise workarounds to minimize the temporal distance among team members (e.g. by meeting synchronously across many different time zones) even at the cost of sacrificing personal time \cite{tang2011your}. 

Noteworthily, the above findings about the causal relations between dispersion and productivity more often come from controlled lab studies rather than investigations of real-world organizations (with valuable exceptions, more on this later).
For studies of real-world teams and organizations, it is often difficult to control or manipulate certain dimensions of team dispersion while controlling others and then compare team outcomes.
\revision{Therefore, instead of real-world settings, researchers often prefer research methods such as controlled lab studies \cite{woolley2010evidence}, ran simulation \cite{zhang2005searching}, or online groups on crowd-sourcing and MOOC platforms}~\cite{wen2017supporting,zhu2012organizing,lykourentzou2016team}. For example, Muller et al.~\cite{muller2014geographical} studied a crowdfunding platform within a company. They measured team outcomes by comparing the funds each crowd-funding team received. They analyzed three dimensions of team dispersion: whether crowd-funding team members are in the same country (a proxy for geographic dispersion), belong to the same department (organizational dispersion), and work for the same globally organized project team (functional dispersion).
They found that teams that shared at least one out of the three ``same-ness'' are more likely to receive crowdfunding investments, and the effects of the three types of ``same-ness'' were approximately additive.
This finding offers valuable empirical backings to Olson and Olson's argument that team dispersion is a multi-dimensional construct~\cite{olson2000dnflustance}; that geographic and temporal distances function together with the socio-demographic, functional, and organizational differences among team members and impact team outcome.
That said, crowd-funding teams can be quite different from real-world project teams, especially in terms of power dynamics and peer relations~\cite{muller2016influences}.

Some prior research did study dispersed teams in real-world organizational contexts. 
McDonald et al. conducted a longitudinal study of co-authoring over distances and studied the relationship between each co-author's perspectives on their shared writing task and their writing outcomes \cite{espinosa2003team}. Grinter, Herbsleb, \& Perry explored geographically distributed software development teams and revealed many approaches the teams had taken to coordinate distributed work \cite{grinter1999geography}.
Overall, this body of work tends to take a qualitative approach. Due to the aforementioned pragmatic reasons, researchers tend to focus on one or a few dimensions of dispersion rather than the "\textit{multi-dimensional construct}". They tend to analyze team outcomes using proxies (e.g. communication effectiveness, influence flow, participation, etc,) rather than direct measures \cite{muller2014geographical,kolari2007structure}.


In this paper, we bridge these two threads of work by examining teams in real-world organizational contexts, using quantitative measures of multiple team dispersion dimensions.


\section{Method}

The goal of this study is to quantitatively examine the relationship between dimensions of team dispersion and team productivity. And we focus on the follow four measurements:

\begin{itemize}
    \item Geographic distance within teams
    \item Temporal distance within teams
    \item Functional distance within teams
    \item Organizational distance within teams
\end{itemize} 

We did not study social-demographic distance because the company considered those personal attributes to be ``sensitive personal information,'' and could not make those attributes available to us for study.

\subsection{Organization Background and Team Overview}
In November 2017, our study site started a research intuitive that pushes for more research work within the company and more research publications at a top-tier engineering research conference. As part of this initiative, 490 researchers, 43 engineers, and 66 managers from offices across the globe \revision{stopped their existing projects, re-organized, and formed 117 new research teams} (597 employees in total; each team on average had 6.68 members, SD=4.88.). 

\rev{These new research project teams were self-organized: any employee could freely write a project proposal or idea and submit it to an executive review committee; once the proposal was approved, members across departments could freely create teams to work on an approved idea; researchers, engineers, and managers could team up together despite they may not usually work together (or form a team without any managers). 
These teams all had the overarching goal of writing up their research projects and submitting them to targeted academic conferences before May 2018, though the specific project goals are up to each team.
Each of the teams was given a Slack group chat channel as the primary communication channel, a Box folder as the primary file sharing space, and an internal Github repository as the primary space for organizing code.
Teams needed to periodically report their results in a short report to the executive committee, and the executive committee had a dashboard to monitor these teams' progress and performance at a scale.
Because this initiative was supposed to encourage cross-team interdisciplinary collaboration, many project teams consist with members from different part of the company.
And depends on the natural of the project idea, some of the teams may need more engineering help, but some others may be primarily carried out by a few researchers.
} 
Eventually, these six-month project teams eventually produced 146 submissions. 79 out of the 117 project teams submitted at least one paper to the designated conference.

\begin{figure}[h!]
        \includegraphics[width=\textwidth]{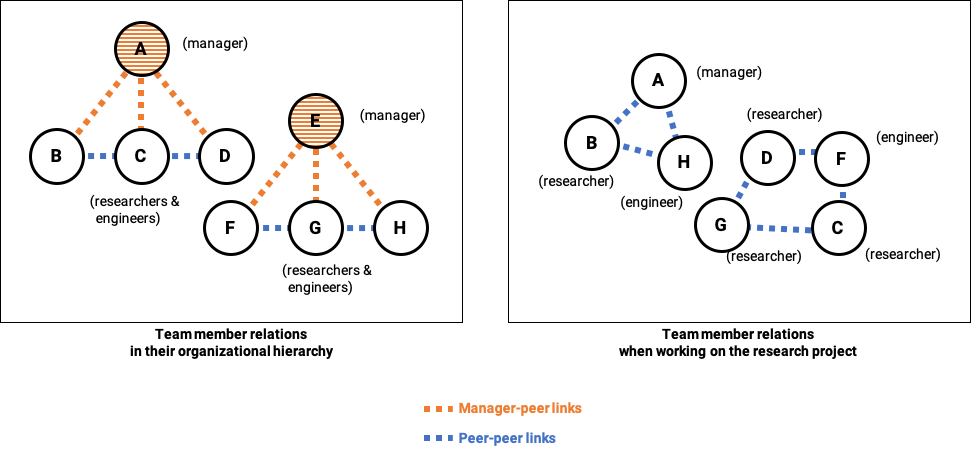}
        \centering
        \captionsetup{width=\textwidth}
        \caption{An illustration of research member relations. During these six-month research projects, team members still belong to their original organizational hierarchy (i.e. They still need to report to their managers on issues not related to the research project). However, when working on the research projects, all members (researchers, managers, engineers) are equal. This setup uniquely allowed us to measure and study organizational dispersion and functional dispersion \textit{separately}.}
        \label{reorg}
\end{figure}

We gained rare access to the teamwork data of these teams. Furthermore, the formation and setup of these teams provided a unique condition for us to study team dispersion and productivity: Team members work from locations across many time zones, allowing us to measure and study temporal and geographical dispersion.
During these six-month research projects, team members still belong to their original organizational hierarchy; They still need to report to their managers on issues not related to the research project. However, when working on the research projects, members of all functional roles (researchers, managers, engineers) are all equal. This setup uniquely allowed us to measure and study organizational dispersion and functional dispersion separately (Figure \ref{reorg}).
Finally, all teams share a clearly defined and observable goal -- i.e. paper submission, allowing us to quantitatively analyze the effects of dispersion on team productivity. Paper submission (rather than paper acceptance) has been a common proxy for measuring research-oriented teamwork~\cite{fox1992research,schiele1995submission}; it provides an observable and comparable measure for creative teams that differ substantially otherwise (in terms of project subject, team progress, measures of success, among others). 

\subsection{Data Collection}
We collected data mainly from three data sources: project team information, organizational hierarchical chart, and paper submission record. Project team information includes team members and project proposals team members collaboratively wrote at the time of team formation. The organizational hierarchical chart includes all employees' job functions (researcher/engineer/manager), office location (office building number, city, country), organizational relationships (i.e. who are their managers and peers).
Paper submission records showed us the number of papers each team submitted.

We aggregated these three datasets using team member information (i.e. email address) only as a unique index key and then \textbf{anonymized} the data before starting data analysis.
\rev{The aggregated dataset is released together with this paper.}

\subsection{Calculating Measures of Team Dispersion}
\label{measures}

Next, we derived team dispersion measures from the data we collected. For each measure, we first computed the distance between each pair of team members:

\begin{itemize}
    \item \textbf{Pairwise geographic dispersion score (geo-distance, integer score)}
     For any given pair of team members, we define their geo-distance as 1 if they are from two different buildings; and 0 if their offices are at the same building. \revision{ This simplification of geo-distance is inspired by the research findings~\cite{kraut1995coordination,allen1984managing} that when two employees work 30 meters apart, their communicate behaviors are similar to the ones working remotely. Our geo-distance variable is to indicate whether they are collocated or remote, thus we design a binary variable.}
     
     \item \textbf{Temporal dispersion score (integer score)}
     We derived the time zone each employee is in from their location data. We then computed temporal distance for each pair of project team members. 
     For example, for a team of three, two of them work from San Francisco (GMT-7) and another from Boston (GMT-4), the pairwise temporal distances between team members are 0, 3, and 3.

    \item \textbf{Pairwise functional dispersion (binary)}
    For each pair of project team members who had the same job role (i.e. both are engineers, researchers, or managers), the functional distance between the two is 0. For those who had different job roles, the functional distance is 1.
    
    \item \textbf{Pairwise organizational distance (integer score)}
    We took a strict hierarchical with peers (SHP) approach \cite{zhang2005searching} when calculating the pair-wise organizational distances among team members. As described in \ref{related-work-measures}, this approach considers both the hierarchical and peer relations in teamwork, and gives equal weights to these relations. We chose this approach rather than a strictly hierarchical one because we consider peer relations important in this particular research teamwork setting: Within each research team, members typically do not work with the managers they usually directly report to (Figure \ref{reorg} right), therefore the peer relations here are equally if not more important than the power/hierarchical distance.
    
\end{itemize}

Next, we computed team dispersion scores by averaging all pair-wise distances among team members. For example, for a team of three members each work from a different building, their pairwise geo-distances are 1, 1, and 1, and the team's geographic dispersion score is 1.
\rev{The use of mean to accumulate group scores is a fairly common practice in various group collaboration studies in CSCW~\cite{olson1993groupwork,yim2017synchronous}.}

In addition to team dispersion characteristics, we also calculated the following descriptive statistics. These measures can additionally impact team performance. We used them in our later data analyses and modeling in order to monitor and eliminate noises.

 \begin{itemize}
\item \textit{Team Size (integer)};
\item \textit{\revision{Team Motivation} (binary)}: During data analysis, \revision{we noticed that in project proposals teams can specify whether or not the publication is their primary team goal (the other option is product).} We suspect this measure can serve as a proxy for team motivation, an important moderator of team productivity. We, therefore, included it in our analysis;
\item \textit{Same Org Team (binary)}: Finally, we analyzed whether each team includes team members exclusively from the same company department. If so, the members are more familiar with each other, which can in turn influence team productivity.

\end{itemize}

\subsection{Statistical Analysis and Modeling}

With all the team dispersion and productivity measures ready, we then started to conduct statistical analysis. \revision{We conducted both a descriptive comparison analysis and a round of regression analysis.} 

\revision{\subsubsection{Comparison Analysis}
We conducted a comparison analysis as a first step to gather some insights for our first research question -- \textbf{\rev{Does each of the quantitative dispersion measures make a difference on team productivity?}}}
To do so, we first divided the 117 research project teams into those \revision{who did and did not produce a paper before the submission deadline} (We will refer to the former as the \textbf{more productive teams} and the latter \textbf{less productive teams}.)
All team-level dispersion scores are continuous independent variables. We therefore first tested the Equality of Variances assumption, then performed independent sample t-tests and Chi-Square tests, comparing the means of dispersion measures between more productive and less productive teams.
For additional categorical independent variables (e.g. team motivation, same org team), we created a two-dimensional table and used the Pearson Chi-Square test to see whether the different groups had disproportionate counts.

\subsubsection{Regression Analysis}
We conducted regression analysis in order to answer our second research question -- \rev{\textbf{Does each of the following measures have an impact on team productivity? }}
First, we checked the assumption for linearity of the logistic and multicollinearity (VIF). We found no interaction effect of independent variables in any of the reported logistic regressions, and we will report VIF next to each model in the Results section.
Next, we generate a logistic regression model for each of the dispersion measures respectively, in order to examine whether and to what extent each measure correlates with the team productivity \revision{with control variables (e.g., Team Size)}. \rev{We chose logistic regression modeling to explore the relationship between each independent variable the dependent variable because our dependent variable is binary (i.e. teams being more productive or less productive), and there are various control variables need to take into account\cite{field2013discovering}}.

Finally, we wanted to compare the predictive power of different dispersion dimensions. To do so, we built a comprehensive full model using all independent variables available, including the four team-dispersion measures and the three additional team characteristics described in \ref{measures}. 

\section{Study Results}
\rev{In total, there are 490 researchers, 43 engineers, and 66 managers from offices across the globe stopped their existing projects, re-organized, and formed 117 new research teams (597 employees in total; each team on average had 6.68 members, SD=4.88.).
The overall statistics are presented in Table \ref{table-team-characteristic-summary}.}

\subsection{The Difference of Team Dispersion Measures between Less Productive and More Productive Teams}

Our analyses revealed that research teams that are more functionally and organizationally disperse were more likely to have successfully produced a research publication in time. \revision{Teams that work across different time zones or geographic locations are not necessarily less (or more) productive.}

\begin{itemize}
    \item Geographical dispersion: More productive and less productive teams are similar in geographical dispersion (t(115)=0.15, p>0.05);
    \item Temporal dispersion: More productive teams have larger temporal dispersion than less productive teams, though the difference is not significant (t(95.98)=-1.46, p>0.05);
    \item Functional dispersion: More productive teams have members of less diverse job roles, in comparison to less productive teams (t(68.20)=2.55, p<0.05);
    \item Organizational dispersion: More productive teams have a significantly higher mean organizational dispersion score (SHP) of 4.12, in comparison to less productive teams which have a mean SHP of 2.98 (t(115)=2.7, p<0.01). In other words, research project teams with members further away from each other in the organizational hierarchy chart were more likely to have successfully produced a publication in time.
\end{itemize}

In addition to dispersion dimensions, we also examined whether other team characteristics influenced productivity. We found that higher team productivity (whether the team eventually produced a publication) is \rev{related to} team motivation (whether the team has highlighted publication as their primary project goal at the time of formation).
We did not observe a difference in team size or team member's original team affiliation between more and less productive teams.
Table \ref{table-team-characteristic-summary} summarizes the results of our comparison analysis. With these insights, we further build regression models to answer our research questions.

\begin{table}[h]
\begin{center}
\begin{tabular}{c|c|c}
\textbf{Team } & \textbf{Less Productive} & \textbf{More Productive} \\ 
\textbf{Characteristics} & \textbf{Teams (N=38)} & \textbf{Teams (N=79)} \\ 
\hline
\rule{0pt}{12pt} Team Motivation** & 12/38 & 46/79 \\
\rule{0pt}{12pt} Average Team Size & 6.66 (SE=1.13) & 6.68 (SE=0.40)\\
\rule{0pt}{12pt} Same Org Team & 4 out of 38 teams (10.53\%) & 5 out of 79 teams (6.31\%) \\
\hline
\rule{0pt}{12pt} Geographic Dispersion Score & 0.32 (SE=0.05) & 0.31 (SE=0.03) \\
\rule{0pt}{12pt} Temporal Dispersion Score & 0.83 (SE=0.26) & 1.36 (SE=0.25) \\
\rule{0pt}{12pt} Functional Dispersion Score** & 0.44 (SE=0.05) & 0.30 (SE=0.03) \\
\rule{0pt}{12pt} \begin{tabular}[c]{@{}c@{}} Organizational Distance Score** \end{tabular} & 2.98 (SE=0.32) & 4.12 (SE=0.25) \\
\end{tabular}
\vspace{10mm}
\
\caption{Team Characteristics and Diversity and Distance Scores Summary in Teams With Submission (N=79) and Teams Without Submissions (N=38). ** indicates difference level is significant p<0.01. Note for continuous variables independent sample t-tests are performed, for discrete variables Chi-Square tests are performed.}
\label{table-team-characteristic-summary}
\end{center}
\end{table}

\subsection{\rev{The Relationship Between Team Dispersion Measures And Productivity}}
\rev{To examine whether and to what extent each team dispersion measure correlate with team productivity outcome}, we built regression models for each dispersion measure.
\rev{The reason why we chose to use regression model instead of a correlation analysis to explore the relationship between each dispersion measure and the outcome is because we want to take the control variables into the equation~\cite{field2013discovering}.
Below we report these regression results to reveal the relationship between each independent variable and our dependent variable.}

\textbf{\textit{\rev{Larger Temporal Dispersion Correlate with Higher Team Productivity}}} 
The Temporal Dispersion Model (Table \ref{table-model-temp}) shows that a higher temporal dispersion score \rev{correlates with} a higher team productivity; team members working from further away time zones were more likely to successfully produce a paper publication.
When the time differences among team members on average increase by one hour, the team is 1.23 times more likely to produce a paper submission. However, \revision{the p-value (0.09)} suggests the relationship is marginal.

\begin{table}[h]
\begin{center}
\begin{tabular}{c|c|c|c|cc}
\textbf{} & \textbf{B} & \textbf{S.E.} & \textbf{p} & \textbf{Exp(B)} & \textbf{VIF} \\
\hline

\rule{0pt}{12pt} \begin{tabular}[c]{@{}c@{}} Temporal Dispersion Score \end{tabular} & 0.21 & 0.12 & 0.09 & 1.23 & 1.03 \\
\rule{0pt}{12pt} \begin{tabular}[c]{@{}c@{}} Team Motivation(Control) \end{tabular} & 1.27 & 0.43 & <.01** & 3.57 & 1.03 \\
\rule{0pt}{12pt} \begin{tabular}[c]{@{}c@{}} \revision{Team Size(Control)} \end{tabular} & -0.02 & 0.04 & 0.71 & 0.98 & 1.03 \\
\rule{0pt}{12pt} \begin{tabular}[c]{@{}c@{}} \revision{Same Org Team(Control)} \end{tabular} & -0.51 & 0.76 & 0.09 & 1.23 & 1.03 \\
\rule{0pt}{12pt} \begin{tabular}[c]{@{}c@{}} Constant \end{tabular} & 0.09 & 0.43 & 0.83 & 1.10 &  \\
\end{tabular}
\vspace{0.3cm}
\captionsetup{width=0.75\textwidth}
\caption{The Temporal Dispersion Model. Higher temporal dispersion score correlates with higher team productivity. $R^2=0.13$}
\label{table-model-temp}
\end{center}
\end{table}

\textbf{\textit{\rev{No evidence of geographical dispersion's correlation with team performance power.}}}
The geographic dispersion model (Table \ref{table-model-geo}) does not reveal meaningful correlation. We found no evidence to support or reject the hypothesis that the geographic dispersion score of a team \rev{correlates with} its productivity.

\begin{table}[h]
\begin{center}
\begin{tabular}{c|c|c|c|cc}
\textbf{} & \textbf{B} & \textbf{S.E.} & \textbf{p} & \textbf{Exp(B)} & \textbf{VIF} \\
\hline
\rule{0pt}{12pt} \begin{tabular}[c]{@{}c@{}} Geographic Dispersion Score \end{tabular} & 0.19 & 0.69 & 0.79 & 1.20 & 1.04 \\
\rule{0pt}{12pt} \begin{tabular}[c]{@{}c@{}} Team Motivation**(Control) \end{tabular} & 1.15 & 0.43 & <.01 & 3.15 & 1.04 \\
\rule{0pt}{12pt} \begin{tabular}[c]{@{}c@{}} \revision{Team Size(Control)} \end{tabular} & -0.00 & 0.05 & 0.95 & 1.00 & 1.04 \\
\rule{0pt}{12pt} \begin{tabular}[c]{@{}c@{}} \revision{Same Org Team(Control)} \end{tabular} & -0.65 & 0.76 & 0.39 & 0.52 & 1.04 \\
\rule{0pt}{12pt} \begin{tabular}[c]{@{}c@{}} Constant \end{tabular} & 0.23 & 0.46 & 0.62 & 1.26 &  \\
\end{tabular}
\vspace{0.3cm}
\captionsetup{width=0.75\textwidth}
\caption{The geographic dispersion model. It shows no evidence to support or reject the geographical dispersion score's relationship to team productivity.$R^2=0.10$}
\label{table-model-geo}
\end{center}
\end{table}

\begin{table}[h]
\begin{center}
\begin{tabular}{c|c|c|c|cc}
\textbf{} & \textbf{B} & \textbf{S.E.} & \textbf{p} & \textbf{Exp(B)} & \textbf{VIF} \\
\hline

\rule{0pt}{12pt} \begin{tabular}[c]{@{}c@{}} Functional Dispersion Score*\end{tabular} & -2.10 & 0.84 & <.05 & 0.12 & 1.04 \\
\rule{0pt}{12pt} \begin{tabular}[c]{@{}c@{}} Team Motivation*(Control) \end{tabular} & 1.00 & 0.43 & <.05 & 2.71 & 1.04 \\
\rule{0pt}{12pt} \begin{tabular}[c]{@{}c@{}} \revision{Team Size(Control)} \end{tabular} & -0.03 & 0.04 & 0.51 & 0.97 & 1.04 \\
\rule{0pt}{12pt} \begin{tabular}[c]{@{}c@{}} \revision{Same Org Team(Control)} \end{tabular} & -1.39 & 0.61 & 0.10 & 0.25 & 1.04 \\
\rule{0pt}{12pt} \begin{tabular}[c]{@{}c@{}} Constant* \end{tabular} & 1.39 & 0.61 & <.05 & 3.99 &  \\
\end{tabular}
\vspace{0.3cm}
\captionsetup{width=0.75\textwidth}
\caption{The functional dispersion model shows that higher functional distance among team members correlates with a team's productivity.$R^2=0.17$}
\label{table-model-func}
\end{center}
\end{table}

\textbf{\textit{\rev{Bigger functional dispersion correlates with a lower team productivity}}} 
The functional dispersion model (Table \ref{table-model-func}) shows that functional distance among team members is correlated with the team's productivity.
\revision{If members of a team have more than one job role, the team is significantly less likely (12\% likelihood) to have a paper submission, in comparison to teams of all the same role (e.g., all researchers). In other words, functionally diverse research teams might be less productive than homogeneous teams.
\rev{Most of the homogeneous teams are researcher-only teams (i.e., no managers, no engineers), so it seems the more researchers in a team, the more likely the team has a paper submission as the outcome.}}

\textbf{\textit{\rev{Organizational dispersion correlates with team productivity.}}}
Our organizational dispersion model (Table \ref{table-model-org}) shows that organizational distance among team members is also \rev{positively correlated with} the team's productivity. 
On average, when a team's organizational dispersion score increases by 1, \revision{the team is 1.36 times more likely to successfully produce a research paper}. 
\rev{To illustrate this result, let us consider this example: Considering a project team that has two researchers, if they are reporting to two different managers and their managers report to the same upper manager, this project team is almost five times more likely to have a paper submission v.s. the case that these two researchers are reporting to the same manager in the same organizational team. }

\begin{table}[h]
\begin{center}
\begin{tabular}{c|c|c|c|cc}
\textbf{} & \textbf{B} & \textbf{S.E.} & \textbf{p} & \textbf{Exp(B)} & \textbf{VIF} \\
\hline

\rule{0pt}{12pt} \begin{tabular}[c]{@{}c@{}} Org Distance Score** (SHP) \end{tabular} & 0.31 & 0.11 & <.01 & 1.36 & 1.00 \\
\rule{0pt}{12pt} \begin{tabular}[c]{@{}c@{}} Team Motivation**(Control) \end{tabular} & 1.21 & 0.44 & <.01 &3.35 & 1.00 \\
\rule{0pt}{12pt} \begin{tabular}[c]{@{}c@{}} \revision{Team Size(Control)} \end{tabular} & -0.03 & 0.05 & 0.51 & 0.97 & 1.00 \\
\rule{0pt}{12pt} \begin{tabular}[c]{@{}c@{}} \revision{Same Org Team(Control)} \end{tabular} & 0.05 & 0.80 & 0.95 & 1.05 & 1.00 \\
\rule{0pt}{12pt} \begin{tabular}[c]{@{}c@{}} Constant \end{tabular} & -0.69 & 0.55 & 0.21 & 0.50 &  \\
\end{tabular}
\vspace{0.3cm}
\captionsetup{width=0.75\textwidth}
\caption{The organizational dispersion model shows that higher organizational distances among team members significantly predicts higher team productivity.$R^2=0.18$}
\label{table-model-org}
\end{center}
\end{table}

\subsection{Comprehensive Full Regression Model}
To compare the predictive power of different dispersion dimensions, we built a ``comprehensive'' regression model.
The resulting model (Table \ref{table-model-all}) suggests that functional dispersion (p=0.05) and team motivation (p<0.05) are the strongest predictors of team productivity. \revision{Organizational dispersion is also predictive, yet did not achieve statistical significance (p=0.10).}

\begin{table}[h]
\begin{center}
\begin{tabular}{c|c|c|c|cc}
\textbf{} & \textbf{B} & \textbf{S.E.} & \textbf{p} & \textbf{Exp(B)} & \textbf{VIF} \\
\hline
\rule{0pt}{12pt} \begin{tabular}[c]{@{}c@{}} \revision{Geographic Dispersion Score} \end{tabular} & -0.86 & 0.84 & 0.31 & 0.42 & 1.04 \\
\rule{0pt}{12pt} \begin{tabular}[c]{@{}c@{}} \revision{Temporal Dispersion Score} \end{tabular} & 0.06 & 0.18 & 0.76 & 1.06 & 1.03 \\
\rule{0pt}{12pt} \begin{tabular}[c]{@{}c@{}} Org Distance Score (SHP) \end{tabular} & 0.28 & 0.17 & 0.10 & 1.33 & 1.05 \\
\rule{0pt}{12pt} \begin{tabular}[c]{@{}c@{}} Functional Dispersion Score \end{tabular} & -1.72 & 0.89 & 0.05 & 0.18 & 1.08 \\
\rule{0pt}{12pt} \begin{tabular}[c]{@{}c@{}} Team Motivation*(Control) \end{tabular} & 1.07 & 0.45 & <.05 &2.92 & 1.04 \\
\rule{0pt}{12pt} \begin{tabular}[c]{@{}c@{}} \revision{Team Size(Control)} \end{tabular} & -0.05 & 0.05 & 0.32 & 0.95 & 1.04 \\
\rule{0pt}{12pt} \begin{tabular}[c]{@{}c@{}} \revision{Same Org Team(Control)} \end{tabular} & -0.64 & 0.90 & 0.48 & 0.53 & 1.04 \\
\rule{0pt}{12pt} \begin{tabular}[c]{@{}c@{}} Constant \end{tabular} & 0.49 & 0.82 & 0.55 & 1.63 &  \\
\end{tabular}
\vspace{0.3cm}
\captionsetup{width=0.75\textwidth}
\caption{The comprehensive full regression model. $R^2=0.23$}
\label{table-model-all}
\end{center}
\end{table}

\section{Discussion}
Geographically dispersed teams are becoming increasingly common in recent years. These teams can face distinctive challenges in communication, coordination, and collaboration, thereby hindering productivity. Understanding these dimensions of team dispersion and their relationship with team productivity is critical for understanding and supporting dispersed teamwork. Extensive prior research has already studied these relations, though often in lab or online settings or using predominantly qualitative measures. This body of work contributed valuable insights into how geographic and temporal distances function together with the socio-demographic, functional, and organizational differences among team members and impact team outcomes.

This paper adds to this body of work by contributing an empirical case study of dispersed teamwork in a real-world organizational setting, using quantitative measures. We studied 117 newly-formed research project teams in a multinational company. During the six months following the team formation, these teams had a shared goal: To produce a research submission to a prestigious conference. 

Interestingly, we found that geographic and time differences are only weakly if at all related to team productivity. However, organizational and functional distances are highly predictive of the productivity of the dispersed teams we studied.
These findings can seem surprising since prior work has more often considered geographic and temporal dispersion rather than organizational distance as major challenges in dispersed teamwork.
In what follows, we first discuss these findings in relation to prior knowledge of dispersed teamwork, providing possible explanations, and identifying open research questions that merit further study. We then discuss implications for remote work technology design. 

\subsection{Towards the Optimal Design of Dispersed Teams}

\subsubsection{Geographical and Temporal Dispersion}

Our results suggest that research teams across multiple time zones are more likely to have higher productivity in paper submission. In view of the well-accepted findings of Olson and Olson~\cite{olson2000dnflustance}, this finding can seem counter-intuitive: We would expect that the more time zone differences in a team, the less productive the team is, as time differences can hinder team coordination and collaboration. 
We speculate that the specific context of the research paper writing task and the unique nature of these research-driven organizational teams could explain this surprising finding. 

While taking all the specific contextual limitations and task uniqueness into consideration, we argue this is still an insightful finding for an organization's remote team formation strategy. Many employees and employers are hesitant to form teams with members coming from different time zones, as such timezone difference imposes lots of challenges on people's schedules and communications~\cite{layton2007software}. However, we found that such a timezone difference can also have benefits. If the organization is a multinational industrial organization, it is inevitable to have people working with others from different time zones. Especially during the COVID-19 like pandemic time, cross-border travel is severely restricted, thus the temporally distributed teams are new for many organizations. Our result can ease the anxiety for these organizations and the team members involved -- at least, such cross-timezone collaboration may help with the team productivity if the team is actively working on paper writing or similar tasks.

\subsubsection{Functional Dispersion}
Prior studies had suggested that ``diversity encourage[s] creativity" \cite{anagnostopoulos2012online, lykourentzou2016personality}; that more diverse teams are better at creative tasks \cite{vasilescu2015gender}.
However, our results seem to contradict this view. Within the research teams we studied, higher functional dispersion significantly predicts lower team productivity. Was it because the research teams were focusing on execution rather than creativity aspects of research, given that they were tasked with a productivity goal (i.e. paper submission) rather than a creative one? Was it because the engineers and managers were less trained in doing research, therefore different teams differ in skill levels? These questions are beyond the scope of our research questions, 
and we propose them for future research.

\subsubsection{Organizational Distance}
Our findings revealed that a higher organizational dispersion score significantly predicts a higher probability that the team will submit a paper.
This result, to some extent, supports the periodical re-organization practice that happens in many organizations. The project teams can benefit from members that are far away from each other on the organizational chart. The employees staying in the same organization team may think too much alike and have overlaps in access to the organizational resources and social capital, thus can be less productive in working on a project. Also, various organizations are experimenting with flatter and more agile ways of organizing project teams. The employee's project team is independent of their organizational team. Our result provides another evidence (further reading ~\cite{amabile2001academic,jarle2011virtual}) for this practice and we argue cross-team project collaboration should be encouraged.

\subsection{\rev{Design Implications for CSCW Systems to Support Team Formation}}

\rev{Our results also have design implications for collaboration support systems at workplace. For example, in our study, we collected three independent data sources and bridged them together to analyze the data. Teams posted their initial project proposal at one place, stored their team member information at a second place, and tracked the project deliverable and productivity at a third place. In the future, organizations may want to build and adopt an inter-operational system(s) to help the organizations and teams to better track project process, thus such system can also provide further opportunity for early intervention, if a team is doomed to failure~\cite{whiting2019did,jung2016coupling}.}

\rev{One thing we noticed in the data collection process of this paper is that organizations nowadays can collect and archive a much richer historical information about the project teams, thanks to the cloud computing technologies. Now that companies have Google Drive or Box to keep track of the artifacts generated from a project team, and who contributed to that artifact; also the code repositories and OverLeaf-like of article drafting systems. The company can leverage these artifacts' meta-data (not to use the content data to avoid privacy concerns), and enable a new future of organizational team collaboration research. But in this paper, we did not have access to those data.}

\rev{The specific findings related to the four types of team dispersion measurements can be incorporated into the new designs of expert recommender systems and team formation systems. For example, the recommender system could suggest team members from farther away organizational team to increase the organizational distance score of a team for certain types of tasks, which may actually have benefits.}

\subsection{Limitations and Future Work}
Our study has a few limitations. First, our study site is one singular multinational company, and the task is academic paper writing. \revision{Even for these research teams, ``delivering a paper'' may not be the best measure to reflect their productivity, as there are multiple steps behind the production of a research paper (e.g., data collection or literature preparation). Thus, some of the findings from this study may or may not apply to other contexts.} We call for more CSCW researchers to follow this thread of work and continue to contribute data points from various other contexts and organizations. Together we can examine the generalizability of the findings from this work.

The individual difference between an employee's research and paper writing skills can be another confounding variable. Especially in our dataset, many of the individuals are engineers and managers, whose primary job role is not research. However, we are looking at the team-level analysis, and some of those factors are already reflected in existing variables, such as the job role diversity score. \revision{Another factor we did not consider is the existing familiarity or relationship between the team members -- some of them may have already worked together with each other while some others may be totally new collaborators. In our log data, we did not have access to those information. In the future, we hope to develop a survey to collect more qualitative and personalized information from individual employees to compliment this study. There are many other similar factors that can be enriched in our study, such as culture background and gender diversity, but due to company's policy, such information are not allowed to be collected or stored in this team project database.}

We acknowledge that the period of this study is pretty short (only six months), but we argue this is a good amount of time for the studied teams to complete a computer science research project and write a paper, and all the studied teams are treated equally, thus we believe the outcomes from these project teams are still comparable. \revision{We collected the data from this fixed time window. This method may not be as well designed as a lab experiment, but it has its strength as it provides a real-world account for the study subjects, as highlighted in ~\cite{russell2014looking,gergle2014experimental}.}

Another limitation is that we used primarily the log data to reflect teams' composition and productivity, and there are other forms of log data that we could leverage in the future (e.g. Slack data~\cite{wang2022group}). But we did not ask employees' perceptions or satisfaction of working in their teams. We acknowledge that the employees' satisfaction is also an important dimension of the outcome of a project, and we plan to run user studies to collect qualitative data to reflect the team relationship and the individual satisfaction perspectives of the team collaboration. Those studies are beyond the scope of this work. We encourage future qualitative research to further examine and improve the quantitative findings from this study, revealing fuller insights into team dispersion and productivity.

At a higher level, explicating the relationship between team dispersion and productivity can inform how we might better design technologies that support dispersed teamwork. For example, the various dimensions of team dispersion scores can be incorporated into the new designs of expert recommender systems and team formation systems. Moreover, team formation and recommendation systems could suggest team members from farther away organizational teams, since increasing the organizational distance score of a team could potentially improve productivity. Examining the findings of this study (especially the findings related to the predictive power of dispersion measures) in the context of Groupware system design should be a critical next step for CSCW research and design.

\section{Conclusion}
In summary, understanding the relationship between team dispersion and productivity is critical for supporting such teams. Extensive prior research has studied these relations in lab settings or using qualitative measures. This paper extends prior work by contributing an empirical case study in a real-world organization, using quantitative measures. We studied 117 new research project teams within a company for 6 months. During this time, all teams shared one goal: submitting a research paper to a conference. We analyzed these teams' dispersion-related characteristics as well as team productivity. Interestingly, we found little statistical evidence that geographic and time differences relate to team productivity. However, organizational distance and functional dispersion are highly predictive of the productivity of the distributed teams we studied. Based on these results, we conclude the paper with socio-technical implications for dispersed organizational team formation strategy and system design as well. We encourage fellow CSCW researchers to further examine these findings in other research settings and to help improve the productivity and quality of distributed teamwork. 

\bibliographystyle{ACM-Reference-Format}
\bibliography{main}


\begin{thebibliography}{60}


\ifx \showCODEN    \undefined \def \showCODEN     #1{\unskip}     \fi
\ifx \showDOI      \undefined \def \showDOI       #1{#1}\fi
\ifx \showISBNx    \undefined \def \showISBNx     #1{\unskip}     \fi
\ifx \showISBNxiii \undefined \def \showISBNxiii  #1{\unskip}     \fi
\ifx \showISSN     \undefined \def \showISSN      #1{\unskip}     \fi
\ifx \showLCCN     \undefined \def \showLCCN      #1{\unskip}     \fi
\ifx \shownote     \undefined \def \shownote      #1{#1}          \fi
\ifx \showarticletitle \undefined \def \showarticletitle #1{#1}   \fi
\ifx \showURL      \undefined \def \showURL       {\relax}        \fi
\providecommand\bibfield[2]{#2}
\providecommand\bibinfo[2]{#2}
\providecommand\natexlab[1]{#1}
\providecommand\showeprint[2][]{arXiv:#2}

\bibitem[\protect\citeauthoryear{Abramo, D’Angelo, and Murgia}{Abramo
  et~al\mbox{.}}{2013}]%
        {abramo2013gender}
\bibfield{author}{\bibinfo{person}{Giovanni Abramo},
  \bibinfo{person}{Ciriaco~Andrea D’Angelo}, {and} \bibinfo{person}{Gianluca
  Murgia}.} \bibinfo{year}{2013}\natexlab{}.
\newblock \showarticletitle{Gender differences in research collaboration}.
\newblock \bibinfo{journal}{\emph{Journal of Informetrics}}
  \bibinfo{volume}{7}, \bibinfo{number}{4} (\bibinfo{year}{2013}),
  \bibinfo{pages}{811--822}.
\newblock


\bibitem[\protect\citeauthoryear{Ahmed and Guha}{Ahmed and Guha}{2012}]%
        {ahmed2012distance}
\bibfield{author}{\bibinfo{person}{Syed~Ishtiaque Ahmed} {and}
  \bibinfo{person}{Shion Guha}.} \bibinfo{year}{2012}\natexlab{}.
\newblock \showarticletitle{Distance matters: an exploratory analysis of the
  linguistic features of Flickr photo tag metadata in relation to impression
  management}. In \bibinfo{booktitle}{\emph{Proceedings of the 2nd ACM SIGMOD
  Workshop on Databases and Social Networks}}. \bibinfo{pages}{7--12}.
\newblock


\bibitem[\protect\citeauthoryear{Allen et~al\mbox{.}}{Allen
  et~al\mbox{.}}{1984}]%
        {allen1984managing}
\bibfield{author}{\bibinfo{person}{Thomas~J Allen} {et~al\mbox{.}}}
  \bibinfo{year}{1984}\natexlab{}.
\newblock \showarticletitle{Managing the flow of technology: Technology
  transfer and the dissemination of technological information within the R\&D
  organization}.
\newblock \bibinfo{journal}{\emph{MIT Press Books}}  \bibinfo{volume}{1}
  (\bibinfo{year}{1984}).
\newblock


\bibitem[\protect\citeauthoryear{Amabile, Patterson, Mueller, Wojcik, Odomirok,
  Marsh, and Kramer}{Amabile et~al\mbox{.}}{2001}]%
        {amabile2001academic}
\bibfield{author}{\bibinfo{person}{Teresa~M Amabile}, \bibinfo{person}{Chelley
  Patterson}, \bibinfo{person}{Jennifer Mueller}, \bibinfo{person}{Tom Wojcik},
  \bibinfo{person}{Paul~W Odomirok}, \bibinfo{person}{Mel Marsh}, {and}
  \bibinfo{person}{Steven~J Kramer}.} \bibinfo{year}{2001}\natexlab{}.
\newblock \showarticletitle{Academic-practitioner collaboration in management
  research: A case of cross-profession collaboration}.
\newblock \bibinfo{journal}{\emph{Academy of Management Journal}}
  \bibinfo{volume}{44}, \bibinfo{number}{2} (\bibinfo{year}{2001}),
  \bibinfo{pages}{418--431}.
\newblock


\bibitem[\protect\citeauthoryear{Anagnostopoulos, Becchetti, Castillo, Gionis,
  and Leonardi}{Anagnostopoulos et~al\mbox{.}}{2012}]%
        {anagnostopoulos2012online}
\bibfield{author}{\bibinfo{person}{Aris Anagnostopoulos}, \bibinfo{person}{Luca
  Becchetti}, \bibinfo{person}{Carlos Castillo}, \bibinfo{person}{Aristides
  Gionis}, {and} \bibinfo{person}{Stefano Leonardi}.}
  \bibinfo{year}{2012}\natexlab{}.
\newblock \showarticletitle{Online team formation in social networks}. In
  \bibinfo{booktitle}{\emph{Proceedings of the 21st international conference on
  World Wide Web}}. ACM, \bibinfo{pages}{839--848}.
\newblock


\bibitem[\protect\citeauthoryear{Bardhan, Krishnan, and Lin}{Bardhan
  et~al\mbox{.}}{2013}]%
        {team-dispersion-Bardhan}
\bibfield{author}{\bibinfo{person}{Indranil Bardhan}, \bibinfo{person}{Vish~V.
  Krishnan}, {and} \bibinfo{person}{Shu Lin}.} \bibinfo{year}{2013}\natexlab{}.
\newblock \showarticletitle{Team Dispersion, Information Technology, and
  Project Performance}.
\newblock \bibinfo{journal}{\emph{Production and Operations Management}}
  \bibinfo{volume}{22}, \bibinfo{number}{6} (\bibinfo{year}{2013}),
  \bibinfo{pages}{1478--1493}.
\newblock
\urldef\tempurl%
\url{https://doi.org/10.1111/j.1937-5956.2012.01366.x}
\showDOI{\tempurl}
\showeprint{https://onlinelibrary.wiley.com/doi/pdf/10.1111/j.1937-5956.2012.01366.x}


\bibitem[\protect\citeauthoryear{Bj{\o}rn, Esbensen, Jensen, and
  Matthiesen}{Bj{\o}rn et~al\mbox{.}}{2014}]%
        {bjorn2014does}
\bibfield{author}{\bibinfo{person}{Pernille Bj{\o}rn}, \bibinfo{person}{Morten
  Esbensen}, \bibinfo{person}{Rasmus~Eskild Jensen}, {and}
  \bibinfo{person}{Stina Matthiesen}.} \bibinfo{year}{2014}\natexlab{}.
\newblock \showarticletitle{Does distance still matter? Revisiting the CSCW
  fundamentals on distributed collaboration}.
\newblock \bibinfo{journal}{\emph{ACM Transactions on Computer-Human
  Interaction (TOCHI)}} \bibinfo{volume}{21}, \bibinfo{number}{5}
  (\bibinfo{year}{2014}), \bibinfo{pages}{1--26}.
\newblock


\bibitem[\protect\citeauthoryear{Bradner and Mark}{Bradner and Mark}{2002}]%
        {bradner2002distance}
\bibfield{author}{\bibinfo{person}{Erin Bradner} {and} \bibinfo{person}{Gloria
  Mark}.} \bibinfo{year}{2002}\natexlab{}.
\newblock \showarticletitle{Why distance matters: effects on cooperation,
  persuasion and deception}. In \bibinfo{booktitle}{\emph{Proceedings of the
  2002 ACM conference on Computer supported cooperative work}}. ACM,
  \bibinfo{pages}{226--235}.
\newblock


\bibitem[\protect\citeauthoryear{Chan, Dow, and Schunn}{Chan
  et~al\mbox{.}}{2014}]%
        {chan2014conceptual}
\bibfield{author}{\bibinfo{person}{Joel Chan}, \bibinfo{person}{Steven Dow},
  {and} \bibinfo{person}{Christian Schunn}.} \bibinfo{year}{2014}\natexlab{}.
\newblock \showarticletitle{Conceptual distance matters when building on
  others' ideas in crowd-collaborative innovation platforms}. In
  \bibinfo{booktitle}{\emph{Proceedings of the companion publication of the
  17th ACM conference on Computer supported cooperative work \& social
  computing}}. \bibinfo{pages}{141--144}.
\newblock


\bibitem[\protect\citeauthoryear{Cramton and Hinds}{Cramton and Hinds}{2004}]%
        {cramton2004subgroup}
\bibfield{author}{\bibinfo{person}{Catherine~Durnell Cramton} {and}
  \bibinfo{person}{Pamela~J Hinds}.} \bibinfo{year}{2004}\natexlab{}.
\newblock \showarticletitle{Subgroup dynamics in internationally distributed
  teams: Ethnocentrism or cross-national learning?}
\newblock \bibinfo{journal}{\emph{Research in organizational behavior}}
  \bibinfo{volume}{26} (\bibinfo{year}{2004}), \bibinfo{pages}{231--263}.
\newblock


\bibitem[\protect\citeauthoryear{Espinosa, Cummings, Wilson, and
  Pearce}{Espinosa et~al\mbox{.}}{2003}]%
        {espinosa2003team}
\bibfield{author}{\bibinfo{person}{J~Alberto Espinosa},
  \bibinfo{person}{Jonathon~N Cummings}, \bibinfo{person}{Jeanne~M Wilson},
  {and} \bibinfo{person}{Brandi~M Pearce}.} \bibinfo{year}{2003}\natexlab{}.
\newblock \showarticletitle{Team boundary issues across multiple global firms}.
\newblock \bibinfo{journal}{\emph{Journal of Management Information Systems}}
  \bibinfo{volume}{19}, \bibinfo{number}{4} (\bibinfo{year}{2003}),
  \bibinfo{pages}{157--190}.
\newblock


\bibitem[\protect\citeauthoryear{Field}{Field}{2013}]%
        {field2013discovering}
\bibfield{author}{\bibinfo{person}{Andy Field}.}
  \bibinfo{year}{2013}\natexlab{}.
\newblock \bibinfo{booktitle}{\emph{Discovering statistics using IBM SPSS
  statistics}}.
\newblock \bibinfo{publisher}{sage}.
\newblock


\bibitem[\protect\citeauthoryear{Fox}{Fox}{1992}]%
        {fox1992research}
\bibfield{author}{\bibinfo{person}{Mary~Frank Fox}.}
  \bibinfo{year}{1992}\natexlab{}.
\newblock \showarticletitle{Research, teaching, and publication productivity:
  Mutuality versus competition in academia}.
\newblock \bibinfo{journal}{\emph{Sociology of education}}
  (\bibinfo{year}{1992}), \bibinfo{pages}{293--305}.
\newblock


\bibitem[\protect\citeauthoryear{Gergle and Tan}{Gergle and Tan}{2014}]%
        {gergle2014experimental}
\bibfield{author}{\bibinfo{person}{Darren Gergle} {and}
  \bibinfo{person}{Desney~S Tan}.} \bibinfo{year}{2014}\natexlab{}.
\newblock \showarticletitle{Experimental research in HCI}.
\newblock In \bibinfo{booktitle}{\emph{Ways of Knowing in HCI}}.
  \bibinfo{publisher}{Springer}, \bibinfo{pages}{191--227}.
\newblock


\bibitem[\protect\citeauthoryear{G{\'o}mez-Zar{\'a}, Paras, Twyman, Lane,
  DeChurch, and Contractor}{G{\'o}mez-Zar{\'a} et~al\mbox{.}}{2019}]%
        {gomez2019would}
\bibfield{author}{\bibinfo{person}{Diego G{\'o}mez-Zar{\'a}},
  \bibinfo{person}{Matthew Paras}, \bibinfo{person}{Marlon Twyman},
  \bibinfo{person}{Jacqueline~N Lane}, \bibinfo{person}{Leslie~A DeChurch},
  {and} \bibinfo{person}{Noshir~S Contractor}.}
  \bibinfo{year}{2019}\natexlab{}.
\newblock \showarticletitle{Who Would You Like to Work With?}. In
  \bibinfo{booktitle}{\emph{Proceedings of the 2019 CHI conference on human
  factors in computing systems}}. \bibinfo{pages}{1--15}.
\newblock


\bibitem[\protect\citeauthoryear{Griffith, Sawyer, and Neale}{Griffith
  et~al\mbox{.}}{2003}]%
        {griffith2003virtualness}
\bibfield{author}{\bibinfo{person}{Terri~L Griffith}, \bibinfo{person}{John~E
  Sawyer}, {and} \bibinfo{person}{Margaret~A Neale}.}
  \bibinfo{year}{2003}\natexlab{}.
\newblock \showarticletitle{Virtualness and knowledge in teams: Managing the
  love triangle of organizations, individuals, and information technology}.
\newblock \bibinfo{journal}{\emph{MIS quarterly}} (\bibinfo{year}{2003}),
  \bibinfo{pages}{265--287}.
\newblock


\bibitem[\protect\citeauthoryear{Grinter, Herbsleb, and Perry}{Grinter
  et~al\mbox{.}}{1999}]%
        {grinter1999geography}
\bibfield{author}{\bibinfo{person}{Rebecca~E Grinter}, \bibinfo{person}{James~D
  Herbsleb}, {and} \bibinfo{person}{Dewayne~E Perry}.}
  \bibinfo{year}{1999}\natexlab{}.
\newblock \showarticletitle{The geography of coordination: dealing with
  distance in R\&D work}. In \bibinfo{booktitle}{\emph{Proceedings of the
  international ACM SIGGROUP conference on Supporting group work}}. ACM,
  \bibinfo{pages}{306--315}.
\newblock


\bibitem[\protect\citeauthoryear{Grudin}{Grudin}{1995}]%
        {grudin1995groupware}
\bibfield{author}{\bibinfo{person}{Jonathan Grudin}.}
  \bibinfo{year}{1995}\natexlab{}.
\newblock \showarticletitle{Groupware and social dynamics: Eight challenges for
  developers}.
\newblock In \bibinfo{booktitle}{\emph{Readings in Human--Computer
  Interaction}}. \bibinfo{publisher}{Elsevier}, \bibinfo{pages}{762--774}.
\newblock


\bibitem[\protect\citeauthoryear{Harris, G{\'o}mez-Zar{\'a}, DeChurch, and
  Contractor}{Harris et~al\mbox{.}}{2019}]%
        {harris2019joining}
\bibfield{author}{\bibinfo{person}{Alexa~M Harris}, \bibinfo{person}{Diego
  G{\'o}mez-Zar{\'a}}, \bibinfo{person}{Leslie~A DeChurch}, {and}
  \bibinfo{person}{Noshir~S Contractor}.} \bibinfo{year}{2019}\natexlab{}.
\newblock \showarticletitle{Joining together online: the trajectory of CSCW
  scholarship on group formation}.
\newblock \bibinfo{journal}{\emph{Proceedings of the ACM on Human-Computer
  Interaction}} \bibinfo{volume}{3}, \bibinfo{number}{CSCW}
  (\bibinfo{year}{2019}), \bibinfo{pages}{1--27}.
\newblock


\bibitem[\protect\citeauthoryear{Harrison and Dourish}{Harrison and
  Dourish}{1996}]%
        {harrison1996re}
\bibfield{author}{\bibinfo{person}{Steve~R Harrison} {and}
  \bibinfo{person}{Paul Dourish}.} \bibinfo{year}{1996}\natexlab{}.
\newblock \showarticletitle{Re-place-ing space: The roles of place and space in
  collaborative systems.}. In \bibinfo{booktitle}{\emph{CSCW}},
  Vol.~\bibinfo{volume}{96}. \bibinfo{pages}{67--76}.
\newblock


\bibitem[\protect\citeauthoryear{Herbsleb, Mockus, Finholt, and
  Grinter}{Herbsleb et~al\mbox{.}}{2000}]%
        {herbsleb2000distance}
\bibfield{author}{\bibinfo{person}{James~D Herbsleb}, \bibinfo{person}{Audris
  Mockus}, \bibinfo{person}{Thomas~A Finholt}, {and} \bibinfo{person}{Rebecca~E
  Grinter}.} \bibinfo{year}{2000}\natexlab{}.
\newblock \showarticletitle{Distance, dependencies, and delay in a global
  collaboration}. In \bibinfo{booktitle}{\emph{Proceedings of the 2000 ACM
  conference on Computer supported cooperative work}}. ACM,
  \bibinfo{pages}{319--328}.
\newblock


\bibitem[\protect\citeauthoryear{Hinds, Retelny, and Cramton}{Hinds
  et~al\mbox{.}}{2015}]%
        {hinds2015flow}
\bibfield{author}{\bibinfo{person}{Pamela Hinds}, \bibinfo{person}{Daniela
  Retelny}, {and} \bibinfo{person}{Catherine Cramton}.}
  \bibinfo{year}{2015}\natexlab{}.
\newblock \showarticletitle{In the flow, being heard, and having opportunities:
  Sources of power and power dynamics in global teams}. In
  \bibinfo{booktitle}{\emph{Proceedings of the 18th ACM Conference on Computer
  Supported Cooperative Work \& Social Computing}}. ACM,
  \bibinfo{pages}{864--875}.
\newblock


\bibitem[\protect\citeauthoryear{Hinds and Bailey}{Hinds and Bailey}{2003}]%
        {hinds2003out}
\bibfield{author}{\bibinfo{person}{Pamela~J Hinds} {and}
  \bibinfo{person}{Diane~E Bailey}.} \bibinfo{year}{2003}\natexlab{}.
\newblock \showarticletitle{Out of sight, out of sync: Understanding conflict
  in distributed teams}.
\newblock \bibinfo{journal}{\emph{Organization science}} \bibinfo{volume}{14},
  \bibinfo{number}{6} (\bibinfo{year}{2003}), \bibinfo{pages}{615--632}.
\newblock


\bibitem[\protect\citeauthoryear{Horwitz and Horwitz}{Horwitz and
  Horwitz}{2007}]%
        {horwitz2007effects}
\bibfield{author}{\bibinfo{person}{Sujin~K Horwitz} {and}
  \bibinfo{person}{Irwin~B Horwitz}.} \bibinfo{year}{2007}\natexlab{}.
\newblock \showarticletitle{The effects of team diversity on team outcomes: A
  meta-analytic review of team demography}.
\newblock \bibinfo{journal}{\emph{Journal of management}} \bibinfo{volume}{33},
  \bibinfo{number}{6} (\bibinfo{year}{2007}), \bibinfo{pages}{987--1015}.
\newblock


\bibitem[\protect\citeauthoryear{Jarle~Gressg{\aa}rd}{Jarle~Gressg{\aa}rd}{2011}]%
        {jarle2011virtual}
\bibfield{author}{\bibinfo{person}{Leif Jarle~Gressg{\aa}rd}.}
  \bibinfo{year}{2011}\natexlab{}.
\newblock \showarticletitle{Virtual team collaboration and innovation in
  organizations}.
\newblock \bibinfo{journal}{\emph{Team Performance Management: An International
  Journal}} \bibinfo{volume}{17}, \bibinfo{number}{1/2} (\bibinfo{year}{2011}),
  \bibinfo{pages}{102--119}.
\newblock


\bibitem[\protect\citeauthoryear{Jung}{Jung}{2016}]%
        {jung2016coupling}
\bibfield{author}{\bibinfo{person}{Malte~F Jung}.}
  \bibinfo{year}{2016}\natexlab{}.
\newblock \showarticletitle{Coupling interactions and performance: Predicting
  team performance from thin slices of conflict}.
\newblock \bibinfo{journal}{\emph{ACM Transactions on Computer-Human
  Interaction (TOCHI)}} \bibinfo{volume}{23}, \bibinfo{number}{3}
  (\bibinfo{year}{2016}), \bibinfo{pages}{1--32}.
\newblock


\bibitem[\protect\citeauthoryear{Kayan, Fussell, and Setlock}{Kayan
  et~al\mbox{.}}{2006}]%
        {kayan2006cultural}
\bibfield{author}{\bibinfo{person}{Shipra Kayan}, \bibinfo{person}{Susan~R
  Fussell}, {and} \bibinfo{person}{Leslie~D Setlock}.}
  \bibinfo{year}{2006}\natexlab{}.
\newblock \showarticletitle{Cultural differences in the use of instant
  messaging in Asia and North America}. In
  \bibinfo{booktitle}{\emph{Proceedings of the 2006 20th anniversary conference
  on Computer supported cooperative work}}. ACM, \bibinfo{pages}{525--528}.
\newblock


\bibitem[\protect\citeauthoryear{Kolari, Finin, Yesha, Yesha, Lyons, Perelgut,
  Hawkins, et~al\mbox{.}}{Kolari et~al\mbox{.}}{2007}]%
        {kolari2007structure}
\bibfield{author}{\bibinfo{person}{Pranam Kolari}, \bibinfo{person}{Tim Finin},
  \bibinfo{person}{Yelena Yesha}, \bibinfo{person}{Yaacov Yesha},
  \bibinfo{person}{Kelly Lyons}, \bibinfo{person}{Stephen Perelgut},
  \bibinfo{person}{Jen Hawkins}, {et~al\mbox{.}}}
  \bibinfo{year}{2007}\natexlab{}.
\newblock \showarticletitle{On the structure, properties and utility of
  internal corporate blogs}. In \bibinfo{booktitle}{\emph{Proceedings of the
  International Conference on Weblogs and Social Media (ICWSM 2007)}}.
\newblock


\bibitem[\protect\citeauthoryear{Kraut and Streeter}{Kraut and
  Streeter}{1995}]%
        {kraut1995coordination}
\bibfield{author}{\bibinfo{person}{Robert~E Kraut} {and}
  \bibinfo{person}{Lynn~A Streeter}.} \bibinfo{year}{1995}\natexlab{}.
\newblock \showarticletitle{Coordination in software development}.
\newblock \bibinfo{journal}{\emph{Commun. ACM}} \bibinfo{volume}{38},
  \bibinfo{number}{3} (\bibinfo{year}{1995}), \bibinfo{pages}{69--82}.
\newblock


\bibitem[\protect\citeauthoryear{Lambiase, Catolino, Tamburri, Serebrenik,
  Palomba, and Ferrucci}{Lambiase et~al\mbox{.}}{2022}]%
        {lambiase2022good}
\bibfield{author}{\bibinfo{person}{Stefano Lambiase}, \bibinfo{person}{Gemma
  Catolino}, \bibinfo{person}{Damian~A Tamburri}, \bibinfo{person}{Alexander
  Serebrenik}, \bibinfo{person}{Fabio Palomba}, {and} \bibinfo{person}{Filomena
  Ferrucci}.} \bibinfo{year}{2022}\natexlab{}.
\newblock \showarticletitle{Good Fences Make Good Neighbours? On the Impact of
  Cultural and Geographical Dispersion on Community Smells}.
\newblock  (\bibinfo{year}{2022}).
\newblock


\bibitem[\protect\citeauthoryear{Layton, Ohland, and Pomeranz}{Layton
  et~al\mbox{.}}{2007}]%
        {layton2007software}
\bibfield{author}{\bibinfo{person}{Richard Layton}, \bibinfo{person}{Matthew
  Ohland}, {and} \bibinfo{person}{Hal Pomeranz}.}
  \bibinfo{year}{2007}\natexlab{}.
\newblock \showarticletitle{Software for student team formation and peer
  evaluation: CATME incorporates Team-Maker}.
\newblock  (\bibinfo{year}{2007}).
\newblock


\bibitem[\protect\citeauthoryear{Lykourentzou, Antoniou, Naudet, and
  Dow}{Lykourentzou et~al\mbox{.}}{2016a}]%
        {lykourentzou2016personality}
\bibfield{author}{\bibinfo{person}{Ioanna Lykourentzou},
  \bibinfo{person}{Angeliki Antoniou}, \bibinfo{person}{Yannick Naudet}, {and}
  \bibinfo{person}{Steven~P Dow}.} \bibinfo{year}{2016}\natexlab{a}.
\newblock \showarticletitle{Personality matters: Balancing for personality
  types leads to better outcomes for crowd teams}. In
  \bibinfo{booktitle}{\emph{Proceedings of the 19th ACM Conference on
  Computer-Supported Cooperative Work \& Social Computing}}. ACM,
  \bibinfo{pages}{260--273}.
\newblock


\bibitem[\protect\citeauthoryear{Lykourentzou, Wang, Kraut, and
  Dow}{Lykourentzou et~al\mbox{.}}{2016b}]%
        {lykourentzou2016team}
\bibfield{author}{\bibinfo{person}{Ioanna Lykourentzou},
  \bibinfo{person}{Shannon Wang}, \bibinfo{person}{Robert~E Kraut}, {and}
  \bibinfo{person}{Steven~P Dow}.} \bibinfo{year}{2016}\natexlab{b}.
\newblock \showarticletitle{Team dating: A self-organized team formation
  strategy for collaborative crowdsourcing}. In
  \bibinfo{booktitle}{\emph{Proceedings of the 2016 CHI Conference Extended
  Abstracts on Human Factors in Computing Systems}}.
  \bibinfo{pages}{1243--1249}.
\newblock


\bibitem[\protect\citeauthoryear{Maznevski and Chudoba}{Maznevski and
  Chudoba}{2000}]%
        {define-distributed-teams-Maznevski2000}
\bibfield{author}{\bibinfo{person}{Martha~L Maznevski} {and}
  \bibinfo{person}{Katherine~M Chudoba}.} \bibinfo{year}{2000}\natexlab{}.
\newblock \showarticletitle{{Bridging Space Over Time: Global Virtual Team
  Dynamics and Effectiveness}}.
\newblock \bibinfo{journal}{\emph{Organization Science}} \bibinfo{volume}{11},
  \bibinfo{number}{5} (\bibinfo{year}{2000}), \bibinfo{pages}{473--492}.
\newblock
\urldef\tempurl%
\url{https://doi.org/10.1287/orsc.11.5.473.15200}
\showDOI{\tempurl}


\bibitem[\protect\citeauthoryear{Mitra, Muller, Shami, Golestani, and
  Masli}{Mitra et~al\mbox{.}}{2017}]%
        {mitra2017spread}
\bibfield{author}{\bibinfo{person}{Tanushree Mitra}, \bibinfo{person}{Michael
  Muller}, \bibinfo{person}{N~Sadat Shami}, \bibinfo{person}{Abbas Golestani},
  {and} \bibinfo{person}{Mikhil Masli}.} \bibinfo{year}{2017}\natexlab{}.
\newblock \showarticletitle{Spread of Employee Engagement in a Large
  Organizational Network: A Longitudinal Analysis}.
\newblock \bibinfo{journal}{\emph{Proceedings of the ACM on Human-Computer
  Interaction}} \bibinfo{volume}{1}, \bibinfo{number}{CSCW}
  (\bibinfo{year}{2017}), \bibinfo{pages}{81}.
\newblock


\bibitem[\protect\citeauthoryear{Muller, Fussell, Gao, Hinds, Oliveira,
  Reinecke, Robert~Jr, Siangliulue, Wulf, and Yuan}{Muller
  et~al\mbox{.}}{2019}]%
        {muller2019learning}
\bibfield{author}{\bibinfo{person}{Michael Muller}, \bibinfo{person}{Susan~R
  Fussell}, \bibinfo{person}{Ge Gao}, \bibinfo{person}{Pamela~J Hinds},
  \bibinfo{person}{Nigini Oliveira}, \bibinfo{person}{Katharina Reinecke},
  \bibinfo{person}{Lionel Robert~Jr}, \bibinfo{person}{Kanya Siangliulue},
  \bibinfo{person}{Volker Wulf}, {and} \bibinfo{person}{Chien-Wen Yuan}.}
  \bibinfo{year}{2019}\natexlab{}.
\newblock \showarticletitle{Learning from Team and Group Diversity: Nurturing
  and Benefiting from our Heterogeneity}. In
  \bibinfo{booktitle}{\emph{Conference Companion Publication of the 2019 on
  Computer Supported Cooperative Work and Social Computing}}.
  \bibinfo{pages}{498--505}.
\newblock


\bibitem[\protect\citeauthoryear{Muller, Geyer, Soule, and Wafer}{Muller
  et~al\mbox{.}}{2014}]%
        {muller2014geographical}
\bibfield{author}{\bibinfo{person}{Michael Muller}, \bibinfo{person}{Werner
  Geyer}, \bibinfo{person}{Todd Soule}, {and} \bibinfo{person}{John Wafer}.}
  \bibinfo{year}{2014}\natexlab{}.
\newblock \showarticletitle{Geographical and organizational distances in
  enterprise crowdfunding}. In \bibinfo{booktitle}{\emph{Proceedings of the
  17th ACM conference on Computer supported cooperative work \& social
  computing}}. ACM, \bibinfo{pages}{778--789}.
\newblock


\bibitem[\protect\citeauthoryear{Muller, Shami, Guha, Masli, Geyer, and
  Wild}{Muller et~al\mbox{.}}{2016}]%
        {muller2016influences}
\bibfield{author}{\bibinfo{person}{Michael Muller}, \bibinfo{person}{N~Sadat
  Shami}, \bibinfo{person}{Shion Guha}, \bibinfo{person}{Mikhil Masli},
  \bibinfo{person}{Werner Geyer}, {and} \bibinfo{person}{Alan Wild}.}
  \bibinfo{year}{2016}\natexlab{}.
\newblock \showarticletitle{Influences of peers, friends, and managers on
  employee engagement}. In \bibinfo{booktitle}{\emph{Proceedings of the 19th
  International Conference on Supporting Group Work}}. ACM,
  \bibinfo{pages}{131--136}.
\newblock


\bibitem[\protect\citeauthoryear{O'Leary and Cummings}{O'Leary and
  Cummings}{2007}]%
        {o2007spatial}
\bibfield{author}{\bibinfo{person}{Michael~Boyer O'Leary} {and}
  \bibinfo{person}{Jonathon~N Cummings}.} \bibinfo{year}{2007}\natexlab{}.
\newblock \showarticletitle{The spatial, temporal, and configurational
  characteristics of geographic dispersion in teams}.
\newblock \bibinfo{journal}{\emph{MIS quarterly}} (\bibinfo{year}{2007}),
  \bibinfo{pages}{433--452}.
\newblock


\bibitem[\protect\citeauthoryear{O'Leary and Mortensen}{O'Leary and
  Mortensen}{2010}]%
        {o2010go}
\bibfield{author}{\bibinfo{person}{Michael~Boyer O'Leary} {and}
  \bibinfo{person}{Mark Mortensen}.} \bibinfo{year}{2010}\natexlab{}.
\newblock \showarticletitle{Go (con) figure: Subgroups, imbalance, and isolates
  in geographically dispersed teams}.
\newblock \bibinfo{journal}{\emph{Organization Science}} \bibinfo{volume}{21},
  \bibinfo{number}{1} (\bibinfo{year}{2010}), \bibinfo{pages}{115--131}.
\newblock


\bibitem[\protect\citeauthoryear{Olson and Olson}{Olson and Olson}{2000}]%
        {olson2000dnflustance}
\bibfield{author}{\bibinfo{person}{Gary~M Olson} {and}
  \bibinfo{person}{Judith~S Olson}.} \bibinfo{year}{2000}\natexlab{}.
\newblock \showarticletitle{Distance matters}.
\newblock \bibinfo{journal}{\emph{Human--computer interaction}}
  \bibinfo{volume}{15}, \bibinfo{number}{2-3} (\bibinfo{year}{2000}),
  \bibinfo{pages}{139--178}.
\newblock


\bibitem[\protect\citeauthoryear{Olson, Olson, Storr{\o}sten, and Carter}{Olson
  et~al\mbox{.}}{1993}]%
        {olson1993groupwork}
\bibfield{author}{\bibinfo{person}{Judith~S Olson}, \bibinfo{person}{Gary~M
  Olson}, \bibinfo{person}{Marianne Storr{\o}sten}, {and} \bibinfo{person}{Mark
  Carter}.} \bibinfo{year}{1993}\natexlab{}.
\newblock \showarticletitle{Groupwork close up: A comparison of the group
  design process with and without a simple group editor}.
\newblock \bibinfo{journal}{\emph{ACM Transactions on Information Systems
  (TOIS)}} \bibinfo{volume}{11}, \bibinfo{number}{4} (\bibinfo{year}{1993}),
  \bibinfo{pages}{321--348}.
\newblock


\bibitem[\protect\citeauthoryear{Olson, Wang, Olson, and Zhang}{Olson
  et~al\mbox{.}}{2017}]%
        {olson2017people}
\bibfield{author}{\bibinfo{person}{Judith~S Olson}, \bibinfo{person}{Dakuo
  Wang}, \bibinfo{person}{Gary~M Olson}, {and} \bibinfo{person}{Jingwen
  Zhang}.} \bibinfo{year}{2017}\natexlab{}.
\newblock \showarticletitle{How people write together now: Beginning the
  investigation with advanced undergraduates in a project course}.
\newblock \bibinfo{journal}{\emph{ACM Transactions on Computer-Human
  Interaction (TOCHI)}} \bibinfo{volume}{24}, \bibinfo{number}{1}
  (\bibinfo{year}{2017}), \bibinfo{pages}{1--40}.
\newblock


\bibitem[\protect\citeauthoryear{Polzer, Crisp, Jarvenpaa, and Kim}{Polzer
  et~al\mbox{.}}{2006}]%
        {Polzer-2006-faultline}
\bibfield{author}{\bibinfo{person}{Jeffrey~T Polzer}, \bibinfo{person}{C~Brad
  Crisp}, \bibinfo{person}{Sirkka~L Jarvenpaa}, {and} \bibinfo{person}{Jerry~W
  Kim}.} \bibinfo{year}{2006}\natexlab{}.
\newblock \showarticletitle{{Extending the Faultline Model to Geographically
  Dispersed Teams: How Colocated Subgroups can Impair Group Functioning}}.
\newblock \bibinfo{journal}{\emph{Academy of Management Journal}}
  \bibinfo{volume}{49}, \bibinfo{number}{4} (\bibinfo{year}{2006}),
  \bibinfo{pages}{679--692}.
\newblock
\urldef\tempurl%
\url{https://doi.org/10.5465/amj.2006.22083024}
\showDOI{\tempurl}


\bibitem[\protect\citeauthoryear{Rogelberg and Rumery}{Rogelberg and
  Rumery}{1996}]%
        {rogelberg1996gender}
\bibfield{author}{\bibinfo{person}{Steven~G Rogelberg} {and}
  \bibinfo{person}{Steven~M Rumery}.} \bibinfo{year}{1996}\natexlab{}.
\newblock \showarticletitle{Gender diversity, team decision quality, time on
  task, and interpersonal cohesion}.
\newblock \bibinfo{journal}{\emph{Small group research}} \bibinfo{volume}{27},
  \bibinfo{number}{1} (\bibinfo{year}{1996}), \bibinfo{pages}{79--90}.
\newblock


\bibitem[\protect\citeauthoryear{Russell and Chi}{Russell and Chi}{2014}]%
        {russell2014looking}
\bibfield{author}{\bibinfo{person}{Daniel~M Russell} {and}
  \bibinfo{person}{Ed~H Chi}.} \bibinfo{year}{2014}\natexlab{}.
\newblock \showarticletitle{Looking back: Retrospective study methods for HCI}.
\newblock In \bibinfo{booktitle}{\emph{Ways of Knowing in HCI}}.
  \bibinfo{publisher}{Springer}, \bibinfo{pages}{373--393}.
\newblock


\bibitem[\protect\citeauthoryear{Schiele}{Schiele}{1995}]%
        {schiele1995submission}
\bibfield{author}{\bibinfo{person}{Jerome~H Schiele}.}
  \bibinfo{year}{1995}\natexlab{}.
\newblock \showarticletitle{Submission rates among African-American faculty:
  The forgotten side of publication productivity}.
\newblock \bibinfo{journal}{\emph{Journal of Social Work Education}}
  \bibinfo{volume}{31}, \bibinfo{number}{1} (\bibinfo{year}{1995}),
  \bibinfo{pages}{46--54}.
\newblock


\bibitem[\protect\citeauthoryear{Tang, Zhao, Cao, and Inkpen}{Tang
  et~al\mbox{.}}{2011}]%
        {tang2011your}
\bibfield{author}{\bibinfo{person}{John~C Tang}, \bibinfo{person}{Chen Zhao},
  \bibinfo{person}{Xiang Cao}, {and} \bibinfo{person}{Kori Inkpen}.}
  \bibinfo{year}{2011}\natexlab{}.
\newblock \showarticletitle{Your time zone or mine?: a study of globally time
  zone-shifted collaboration}. In \bibinfo{booktitle}{\emph{Proceedings of the
  ACM 2011 conference on Computer supported cooperative work}}. ACM,
  \bibinfo{pages}{235--244}.
\newblock


\bibitem[\protect\citeauthoryear{Van~Knippenberg and Schippers}{Van~Knippenberg
  and Schippers}{2007}]%
        {van2007work}
\bibfield{author}{\bibinfo{person}{Daan Van~Knippenberg} {and}
  \bibinfo{person}{Michaela~C Schippers}.} \bibinfo{year}{2007}\natexlab{}.
\newblock \showarticletitle{Work group diversity}.
\newblock \bibinfo{journal}{\emph{Annu. Rev. Psychol.}}  \bibinfo{volume}{58}
  (\bibinfo{year}{2007}), \bibinfo{pages}{515--541}.
\newblock


\bibitem[\protect\citeauthoryear{Vasilescu, Posnett, Ray, van~den Brand,
  Serebrenik, Devanbu, and Filkov}{Vasilescu et~al\mbox{.}}{2015}]%
        {vasilescu2015gender}
\bibfield{author}{\bibinfo{person}{Bogdan Vasilescu}, \bibinfo{person}{Daryl
  Posnett}, \bibinfo{person}{Baishakhi Ray}, \bibinfo{person}{Mark~GJ van~den
  Brand}, \bibinfo{person}{Alexander Serebrenik}, \bibinfo{person}{Premkumar
  Devanbu}, {and} \bibinfo{person}{Vladimir Filkov}.}
  \bibinfo{year}{2015}\natexlab{}.
\newblock \showarticletitle{Gender and tenure diversity in GitHub teams}. In
  \bibinfo{booktitle}{\emph{Proceedings of the 33rd annual ACM conference on
  human factors in computing systems}}. \bibinfo{pages}{3789--3798}.
\newblock


\bibitem[\protect\citeauthoryear{Wang, Olson, Zhang, Nguyen, and Olson}{Wang
  et~al\mbox{.}}{2015}]%
        {wang2015docuviz}
\bibfield{author}{\bibinfo{person}{Dakuo Wang}, \bibinfo{person}{Judith~S
  Olson}, \bibinfo{person}{Jingwen Zhang}, \bibinfo{person}{Trung Nguyen},
  {and} \bibinfo{person}{Gary~M Olson}.} \bibinfo{year}{2015}\natexlab{}.
\newblock \showarticletitle{DocuViz: visualizing collaborative writing}. In
  \bibinfo{booktitle}{\emph{Proceedings of the 33rd Annual ACM conference on
  human factors in computing systems}}. \bibinfo{pages}{1865--1874}.
\newblock


\bibitem[\protect\citeauthoryear{Wang, Wang, Yu, Ashktorab, and Tan}{Wang
  et~al\mbox{.}}{2022}]%
        {wang2022group}
\bibfield{author}{\bibinfo{person}{Dakuo Wang}, \bibinfo{person}{Haoyu Wang},
  \bibinfo{person}{Mo Yu}, \bibinfo{person}{Zahra Ashktorab}, {and}
  \bibinfo{person}{Ming Tan}.} \bibinfo{year}{2022}\natexlab{}.
\newblock \showarticletitle{Group Chat Ecology in Enterprise Instant Messaging:
  How Employees Collaborate Through Multi-User Chat Channels on Slack}.
\newblock \bibinfo{journal}{\emph{Proceedings of the ACM on Human-Computer
  Interaction}} \bibinfo{volume}{6}, \bibinfo{number}{CSCW1}
  (\bibinfo{year}{2022}), \bibinfo{pages}{1--14}.
\newblock


\bibitem[\protect\citeauthoryear{Wen, Maki, Dow, Herbsleb, and Rose}{Wen
  et~al\mbox{.}}{2017}]%
        {wen2017supporting}
\bibfield{author}{\bibinfo{person}{Miaomiao Wen}, \bibinfo{person}{Keith Maki},
  \bibinfo{person}{Steven Dow}, \bibinfo{person}{James~D Herbsleb}, {and}
  \bibinfo{person}{Carolyn Rose}.} \bibinfo{year}{2017}\natexlab{}.
\newblock \showarticletitle{Supporting virtual team formation through
  community-wide deliberation}.
\newblock \bibinfo{journal}{\emph{Proceedings of the ACM on Human-Computer
  Interaction}} \bibinfo{volume}{1}, \bibinfo{number}{CSCW}
  (\bibinfo{year}{2017}), \bibinfo{pages}{109}.
\newblock


\bibitem[\protect\citeauthoryear{WHITING, BLAISING, BARREAU, FIUZA, MARDA,
  VALENTINE, and BERNSTEIN}{WHITING et~al\mbox{.}}{2019}]%
        {whiting2019did}
\bibfield{author}{\bibinfo{person}{MARK~E WHITING}, \bibinfo{person}{ALLIE
  BLAISING}, \bibinfo{person}{CHLOE BARREAU}, \bibinfo{person}{LAURA FIUZA},
  \bibinfo{person}{NIK MARDA}, \bibinfo{person}{MELISSA VALENTINE}, {and}
  \bibinfo{person}{MICHAEL~S BERNSTEIN}.} \bibinfo{year}{2019}\natexlab{}.
\newblock \showarticletitle{Did It Have To End This Way? Understanding the
  Consistency of Team Fracture}.
\newblock  (\bibinfo{year}{2019}).
\newblock


\bibitem[\protect\citeauthoryear{Williams and O’Reilly~III}{Williams and
  O’Reilly~III}{1998}]%
        {williams1998demography}
\bibfield{author}{\bibinfo{person}{KY Williams} {and} \bibinfo{person}{CA
  O’Reilly~III}.} \bibinfo{year}{1998}\natexlab{}.
\newblock \showarticletitle{Demography and Diversity in Organisations: A review
  of 40 years of research in BM Staw and LL Cummings (eds) Research in
  Organisational Behaviour Vol. 20}.
\newblock \bibinfo{journal}{\emph{Jai Pres, Connecticut}}
  (\bibinfo{year}{1998}).
\newblock


\bibitem[\protect\citeauthoryear{Woolley, Chabris, Pentland, Hashmi, and
  Malone}{Woolley et~al\mbox{.}}{2010}]%
        {woolley2010evidence}
\bibfield{author}{\bibinfo{person}{Anita~Williams Woolley},
  \bibinfo{person}{Christopher~F Chabris}, \bibinfo{person}{Alex Pentland},
  \bibinfo{person}{Nada Hashmi}, {and} \bibinfo{person}{Thomas~W Malone}.}
  \bibinfo{year}{2010}\natexlab{}.
\newblock \showarticletitle{Evidence for a collective intelligence factor in
  the performance of human groups}.
\newblock \bibinfo{journal}{\emph{science}} \bibinfo{volume}{330},
  \bibinfo{number}{6004} (\bibinfo{year}{2010}), \bibinfo{pages}{686--688}.
\newblock


\bibitem[\protect\citeauthoryear{Yim, Wang, Olson, Vu, and Warschauer}{Yim
  et~al\mbox{.}}{2017}]%
        {yim2017synchronous}
\bibfield{author}{\bibinfo{person}{Soobin Yim}, \bibinfo{person}{Dakuo Wang},
  \bibinfo{person}{Judith Olson}, \bibinfo{person}{Viet Vu}, {and}
  \bibinfo{person}{Mark Warschauer}.} \bibinfo{year}{2017}\natexlab{}.
\newblock \showarticletitle{Synchronous writing in the classroom:
  Undergraduates’ collaborative practices and their impact on text quality,
  quantity, and style}. In \bibinfo{booktitle}{\emph{Proceedings of the
  Conference on Computer Supported Cooperative Work (CSCW’17)}},
  Vol.~\bibinfo{volume}{10}.
\newblock


\bibitem[\protect\citeauthoryear{Zhang and Ackerman}{Zhang and
  Ackerman}{2005}]%
        {zhang2005searching}
\bibfield{author}{\bibinfo{person}{Jun Zhang} {and} \bibinfo{person}{Mark~S
  Ackerman}.} \bibinfo{year}{2005}\natexlab{}.
\newblock \showarticletitle{Searching for expertise in social networks: a
  simulation of potential strategies}. In \bibinfo{booktitle}{\emph{Proceedings
  of the 2005 international ACM SIGGROUP conference on Supporting group work}}.
  ACM, \bibinfo{pages}{71--80}.
\newblock


\bibitem[\protect\citeauthoryear{Zheng, Vogelsang, and Pinkwart}{Zheng
  et~al\mbox{.}}{2015}]%
        {zheng2015impact}
\bibfield{author}{\bibinfo{person}{Zhilin Zheng}, \bibinfo{person}{Tim
  Vogelsang}, {and} \bibinfo{person}{Niels Pinkwart}.}
  \bibinfo{year}{2015}\natexlab{}.
\newblock \showarticletitle{The impact of small learning group composition on
  student engagement and success in a MOOC}. In
  \bibinfo{booktitle}{\emph{Proceedings of the 8th International Conference of
  Educational Data Mining}}. \bibinfo{pages}{500--503}.
\newblock


\bibitem[\protect\citeauthoryear{Zhu, Kraut, and Kittur}{Zhu
  et~al\mbox{.}}{2012}]%
        {zhu2012organizing}
\bibfield{author}{\bibinfo{person}{Haiyi Zhu}, \bibinfo{person}{Robert Kraut},
  {and} \bibinfo{person}{Aniket Kittur}.} \bibinfo{year}{2012}\natexlab{}.
\newblock \showarticletitle{Organizing without formal organization: group
  identification, goal setting and social modeling in directing online
  production}. In \bibinfo{booktitle}{\emph{Proceedings of the ACM 2012
  conference on Computer Supported Cooperative Work}}.
  \bibinfo{pages}{935--944}.
\newblock


\end{thebibliography}

\received{January 2022}
\received[revised]{April 2022}
\received[accepted]{May 2022}

\end{document}